\RequirePackage{fix-cm}
\documentclass[twocolumn]{svjour3}          
\smartqed  
\usepackage{graphicx}
\journalname{Eur. Phys. J. D}
\begin{document}

\title{{Nanoclusters and nanoscale voids as possible sources of increasing dark current in high-gradient vacuum breakdown}}


\author{I.I. Musiienko         \and
        S.O. Lebedynskyi \and 
        R.I. Kholodov
}


\institute{
              Institute of Applied Physics, National Academy of Sciences of Ukraine, Petropavlivska str. 58, 40000 Sumy, Ukraine \\
              Tel.: +380-542-222-794\\
              Fax: +380-542-223-760\\
              \email{igor-musienko@ukr.net}           
         }

\date{Received: date / Accepted: date}

\maketitle

\begin{abstract}
The potential barrier model considers an additional current that can lead to the high-gradient breakdowns in accelerating structures is proposed. An oscillatory resonance feature of the field emission current from a double-layer metal system with a nanoscale coating is shown. The double  potential barrier was used for calculations of the field emission current density value. The presence of resonant properties of the field emission current density of a double-layer metal system with a nanometric coating is revealed. The field emission current increasing more than 5 times (for the constant value of an electric field strength $E~=~5 ~GV~/~m$) when considering not ideal copper surface with presence of nanoclusters on the surface of or nanoscale voids in the near-surface layer is observed. 
\end{abstract}

\section{Introduction}
\label{intro}

Field emission is a type of emission that is caused by the quantum mechanical tunneling of electrons from the metal surface to a vacuum induced by an electrostatic field \cite{r1}. This phenomenon is typical for strong fields with electric field strength $E = 10^{8}\div 10^{10}$ ~V~/~m. It is one of the main factors, which leads to the dark currents occurring in the accelerating structures and, as a consequence, the loss of electrical insulating properties of the interelectrode gap \cite{r2,r3,r4}. The problem of regulating the magnitude of the field emission current is relevant, in particular, to preventing high-vacuum high-gradient breakdowns. Another important goal is to obtain gradient-stable materials that eliminate the possibility of electric discharge in the components of modern accelerators. The amplification of the field emission current is also important, particularly for the operation of field emission sources, tunnel microscopy requirements, electron holography, vacuum nanoelectronics \cite{r5,r6,r7,r8,r9}.

\begin{figure}
	\includegraphics[width=0.5\textwidth]{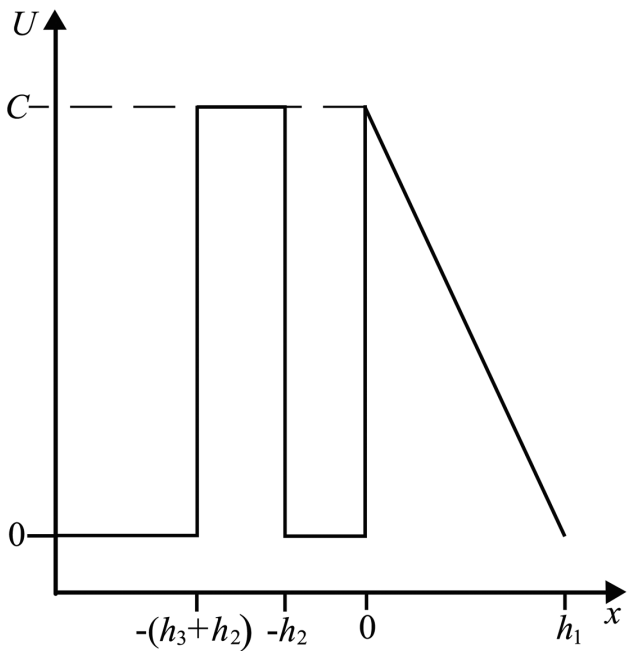}
	\caption{Double potential barrier model of rectangular-triangular shape}
	\label{penA}       
\end{figure}

Modeling of experimental operating conditions of the compact linear electron-positron collider in the CLIC (Compact Linear Collider) project is carried out at CERN (European Organization for Nuclear Research). It was showed when applied a high-frequency electromagnetic field to the metal surfaces of accelerating structures, which provides an acceleration rate of the order of 100 ~MV~/~m, the high-gradient high-vacuum breakdowns occurs \cite{r2,r3,r4}. Since the breakdowns leads to a significant loss of energy and to the damaging surface of facilities components, reducing the probability of its occurrence in the CLIC accelerating structures is a necessary condition for the high-quality functioning of the future collider.

Therefore, various methods of increasing stability of the accelerator modules surface to high-gradient high-vacuum breakdown are being studied. Among the possible methods to prevent high-gradient breakdowns are metal surface modification by implanting argon, nitrogen, zirconium ions into the near-surface copper layer (which is the base material of CLIC accelerating structure disks) and an ion-plasma treatment of the substrate surface by a metal layer with a greater work function \cite{r10}.

The effect of changing the electrode surface characteristics on the field emission current density was experimentally investigated in \cite{r10} and theoretically in \cite{r11,r12}. In the article \cite{r10}, it was experimentally confirmed that the condition of the electrode surface significantly affects the breakdown voltage. Plasma and ion-beam modification of the copper surface leads to an increase in the high-vacuum breakdown voltage from 5 to 35~\% and decrease the dark current in the interelectrode gap. The result with the greatest suppression of the high-vacuum breakdown in the experiments was obtained for copper with the TiN coating and additional irradiation of the surface with the argon ions energy is equal to 300 ~keV.

In articles \cite{r13,r14} an effect of a magnetic field on the field emission current was interpreted. The breakdown voltage decrease due to an increase of ionization by the electron impact was predicted. In \cite{r15} the possibility of considering the relativistic correction for the field emission current density of electrons from metal is described theoretically. It was found that in the case of the field electron emission from pulsars the relativistic change in the current value is about 10~\% of the Fowler-Nordheim current density. In \cite{r16} it was found that the presence of an external magnetic field parallel to the metal surface with a value of 10 ~T leads to a decrease in the transmission coefficient by 10~\%. In paper \cite{r17} the flash emission with a duration of 1~-~5~ns, induced by a pulsed electric field $E \ge 10^{7}$ ~V~/~m applied to a cathode, produced of graphene-like structures, was experimentally detected. In contrast to explosive emission \cite{r7} the new type of emission is not associated with the appearance of plasma and destruction of local nano- and microregions of the cathode surface.

The investigation \cite{r18} shows an effect of the Coulomb blockade phenomenon on the current density value from nanoscale objects. In \cite{r19} the low-voltage field emission of electrons from nanostructured carbon materials is theoretically described due to the presence of resonant surface states, provided that the emitter surface has two carbon phases with different electronic properties. The possibility of increasing the emission current due to the resonant tunnelling of electrons through the double potential barrier by four or more orders of magnitude was predicted.

Investigations of the vacuum breakdown mechanism at the prototypes of the future CLIC accelerator showed that the dark current values cannot be explained only by an electric field enhancement by metal surface irregularities \cite{r20}. The field emission measurements on a macroscopic system needs a large – 30, 100 or more – correction factor to get current $I$ vs voltage $U$ data, corresponding to the current density value from the Fowler-Nordheim equation. However, an amplification factor is not enough for corrections of 30 to 100 and more in the case of measuring the field emission current density from a copper surface in experiments performed in the CERN DC Spark Systems (electrodes in ultra high vacuum with a high voltage external circuit, used to study breakdown). Thus, possible sources of dark current amplification, which leads to high-vacuum breakdowns, are now being searched for.

The main idea of this paper is to study the possibility of increasing the field electron emission current in the case of two-layer metal systems by choosing the coating thickness on the substrate. It is similar to a special technique of an anti-reflective coating to reduce losses for the reflection of light \cite{r21,r22,r23}. Figure~\ref{penA} shows the shape of the double potential barrier, which is considered for calculating the field electron emission current from a two-layer metal system. Similar models of the potential barrier were used to study the properties of quantum wells and resonant tunnelling of the field electron emission through the double barrier \cite{r24,r25,r26,r27,r28}.

The paper \cite{r29}  presents an exact analytical theory for field emission from dielectric coated cathode surfaces, by solving the one-dimensional (1D) Schr$\ddot{o}$dinger equation with a double-triangular potential barrier, which is formed by applying dc electric field to the dielectric coated cathode surface. The quantum model in \cite{r29} is also compared with a modified Fowler-Nordheim equation for a double-triangular barrier, showing qualitatively good agreement.

The purpose of this study is to perform a numerical calculation of the field electron emission current using the model of the double potential barrier of a rectangular-triangular shape. One of the issues is to explain the analytical relationship between the de Broglie wavelength of the tunnelling electron and the width of the potential well in the case of the double rectangular potential barrier. An important issue is to demonstrate the effect of resonant tunnelling on the field emission current density due to the double potential barrier. This theory is used to find additional sources of increasing the field emission current from the metal surface.

\section{Replacement of a single-stage triangular potential barrier with a rectangular one of equal transparency}

The double potential barrier for barrier structures in semiconductor layered systems is considered in reference \cite{r24}. It is also used to evaluate the value of the field emission current from a double-layer metal-semiconductor system in \cite{r29,r30} and for estimates the current density from the near-surface copper layer modiﬁed by the point vacancies in \cite{r12}. The proposed model makes it possible to describe two different objects such as nanoclusters and nanoscale voids.

If there is a nanoscale void in the surface metal layer, in Figure~\ref{penA} the model represents $h_1$ as the barrier width at the metal-vacuum interface, $h_2$ is the distance from the metal-vacuum interface to the location of the nanoscale void in the surface metal layer, $h_3$ is the nanoscale void diameter. If the nanocluster is located on the metal surface, then the barrier parameters in Figure~\ref{penA} can be explained as $h_1$ is the barrier width at the metal-vacuum interface, $h_2$ is the nanocluster size, $h_3$ is the dipole layer thickness of the metal to metal contact. Below in this study, the authors will consider the nanoscale voids.

\begin{figure}
	\includegraphics[width=0.5\textwidth]{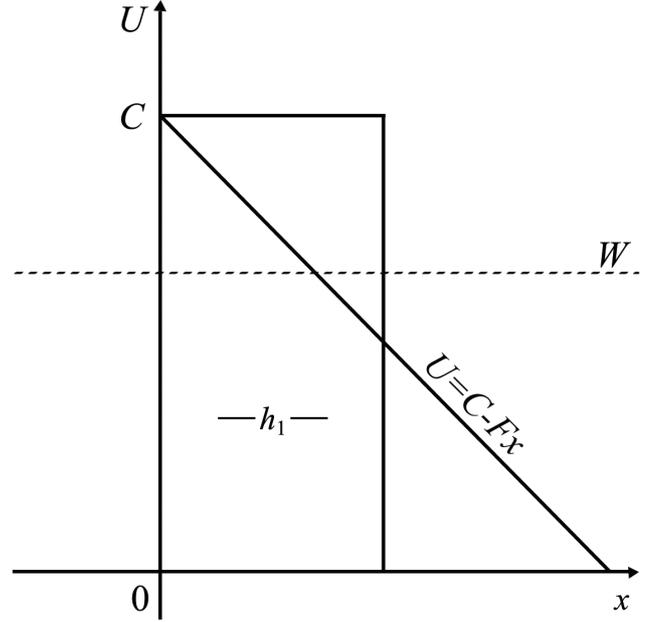}
	\caption{Replacement of the triangular shape of the
		potential barrier with a rectangular one, provided
		that their transmission coefficients are equal}
	\label{penB}       
\end{figure}

Note that for an authentic model of the potential barrier, it is necessary to take into account the presence of electric field strength in the dipole layer, which occurs upon contact of surfaces of various metals \cite{r30}. In the case of two layers of identical metal, the dipole dielectric field is zero.

Fowler and Nordheim found the formula for the current emission density $j_{F-N}$ based on the determination of the transmission coefficient $D_{F-N}$ as a function of the electron energy $W$ in the case of the triangular potential barrier at the boundary from the metal to a vacuum. In this investigation, it will be replaced the authentic triangular potential barrier shape shown in Figure~\ref{penB} by the rectangular one provided that their transmission coefficients are equal. Thus, it is possible to significantly simplify and speed up numerical calculations and avoid mathematical transformations with Hankel functions of the second kind to find the transmission coefficient of the potential barrier. Let us introduce the notation: $U$ is the potential energy; the potential energy height $C=\mu+\chi$; $\mu$ is the chemical potential; $\chi$ is the work function; $F = e\cdot{E}$; $e$ is the electron charge; $E$ is the electric field strength.

The transmission coefficient of a rectangular potential barrier of width $h_1$ \cite{r24,r31} can be written in the following analytical form in selected convenient notations:

\begin{equation}
	D_{1}  = \left(1+\frac{\eta^{2}}{4}sh^{2}(k_{2}h_{1})\right)^{-1},
	\label{as1}
\end{equation}
where $\eta=\frac{C}{\sqrt{W\varphi}}$; $k_{2}=k\sqrt{\varphi}$; $k=\frac{\sqrt{2m_{e}}}{\hbar}$; $m_{e}$ is the electron mass; $\varphi=C-W$.

The expression for the transmission coefficient $D_{F-N}$ of the triangular potential barrier, obtained by Fowler and Nordheim \cite{r1}, is necessary to find the field electron emission current density at a range of values of the electric field strength $E = 10^{8}\div 10^{10}$ ~V~/~m and is equal to:
\begin{equation}
	D_{F-N}  =\frac{4}{\eta}\exp{\left(-\frac{4k\varphi^{\frac{3}{2}}}{3F}\right)}.
	\label{as2}
\end{equation}
For $F>10^{10}$ ~eV~/~m it is necessary to use the numerical method of obtaining the complex amplitudes of wave functions and the transmission coefficient of the potential barrier because formula (\ref{as2}) for the above interval is not accurate.
From the equality $D_{1}=D_{F-N}$ it was obtained the dependence of $h_{1}$ on the physical quantities $W$ and $F$ in general form, which is correct for the above-mentioned range of values of the electric field strength $E$:
\begin{eqnarray}
	h_{1}  = \frac{1}{k_{2}}\times \nonumber\\
	\times {arsh\left(\frac{2}{C}\cdot\sqrt{\frac{W}{D_{F-N}}\left(D_{F-N}W-D_{F-N}C+\varphi\right)}\right)},
	\label{as3}
\end{eqnarray}
where $arsh(x)$ is an inverse hyperbolic sine. On condition $k_{2}h_{1}\gg1$, equating the right-hand sides of formulas (\ref{as1}) and (\ref{as2}) and not taking into consideration the values of the factors before the exponents of the expressions, an approximate formula was found for the effective $h_{1eff}$:
\begin{equation}
	h_{1eff} \approx \frac{2\varphi}{3F}.
	\label{as4}
\end{equation}
The approximate expression to formula (\ref{as3}) can be written as:
\begin{equation}
	h_{1} \approx \frac{2\varphi}{3F}+\frac{1}{2k_{2}}\ln{\left(\frac{4}{\eta}\right)}.
	\label{as5}
\end{equation}

\begin{figure}
	\includegraphics[width=0.5\textwidth]{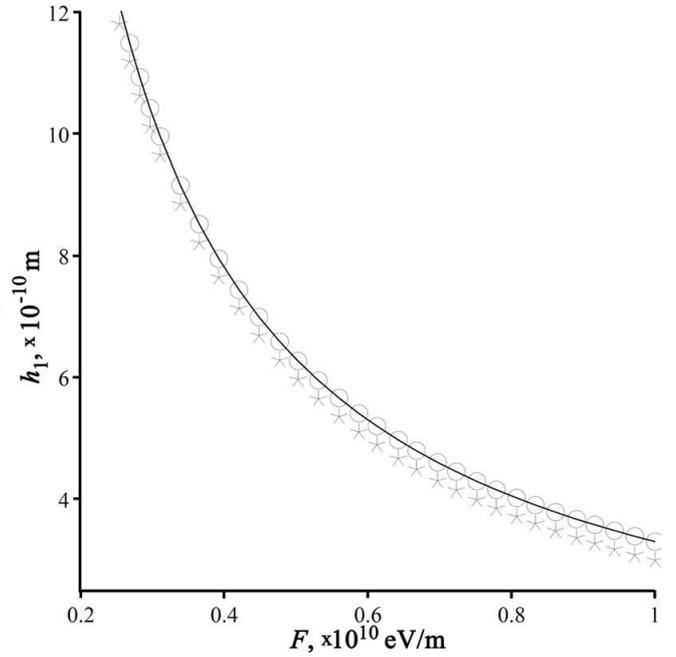}
	\caption{Graphs of dependence $h_{1}(F)$: formula (\ref{as3}) is represented 
		by a solid line; expression (\ref{as4}) is shown by a line of asterisks; 
		formula (\ref{as5}) is depicted by a line of rings}
	\label{penC}       
\end{figure}

Figure~\ref{penC} shows the graphical dependences of $h_{1}$ on the value of $F$ for the parameters of copper $C = 12$ ~eV, $W = \mu$. For calculations, it is advisable to take the electron energy near the energy value of the chemical potential $\mu$, as the most probable current carriers of the field emission from the metal surface are free electrons at the upper energy levels according to Sommerfeld's theory \cite{r1,r8}. The graphs of dependences (\ref{as3}) (indicated in Figure~\ref{penC} by a solid line) and (\ref{as5}) (a line of rings) coincide on the considered interval of values of $F$. Formula (\ref{as4}) (the line is represented by asterisks) approximately describes the analytical dependence $h_{1}(F)$.
\begin{figure}
	\includegraphics[width=0.5\textwidth]{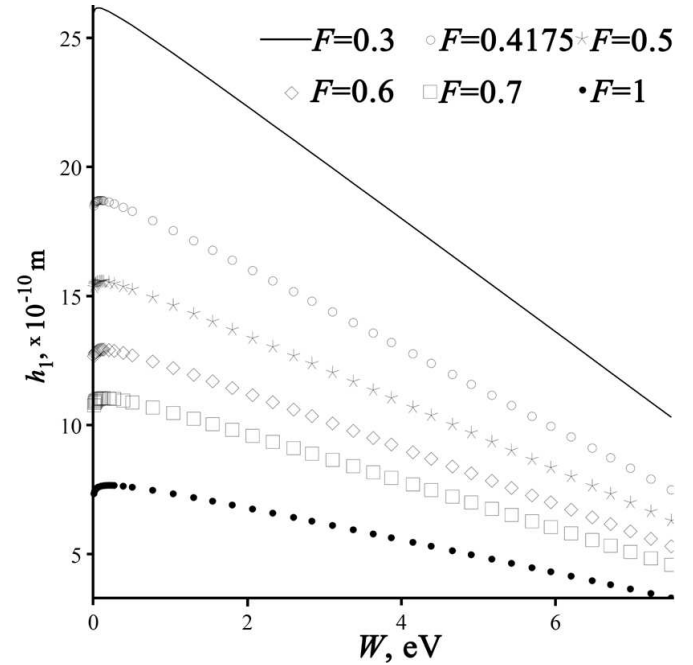}
	\caption{Graphic dependencies $h_{1}(W)$}
	\label{penD}       
\end{figure}

\begin{figure}
	\includegraphics[width=0.5\textwidth]{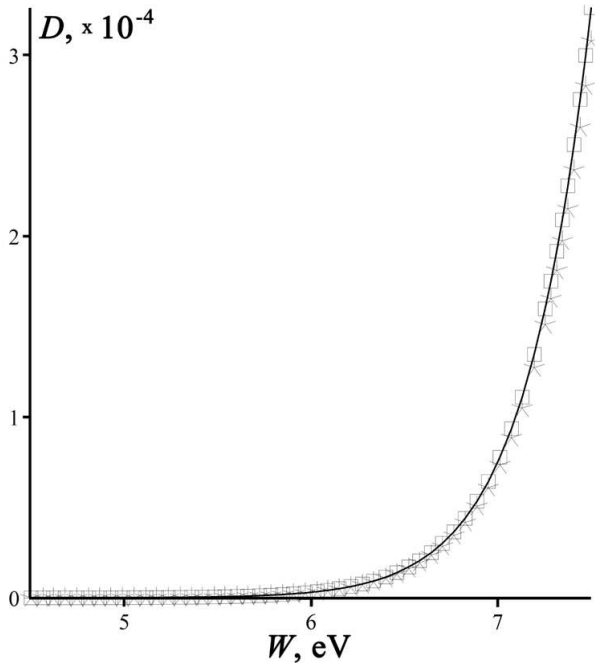}
	\caption{Graphic dependences $D(W)$ in the case of single-stage potential barriers}
	\label{penE}       
\end{figure}

Figure~\ref{penD} demonstrates the dependences of the width $h_{1}$ of a rectangular potential barrier on the electron energy for different values of $F$. Thus, if the constant electric field strength increases, then the value of $h_{1}$ decreases. As a consequence, the width of the rectangular potential barrier narrows, and for values of $F>10^{10}$ ~eV~/~m, the transmission coefficient $D_{1}$ increases significantly.

Figure~\ref{penE} represents the graphic dependences of the transmission coefficient $D$ of single-stage potential barriers of triangular and rectangular shapes on the electron energy for the parameters: $C = 12$ ~eV, $\mu = 7.5$ ~eV, $F = 7.5\cdot10^{9}$ ~eV~/~m. The quantity $h_{1}$ is determined for an arbitrary energy $W$ using expression (\ref{as5}). The curves of dependences are indicated by a solid line in the case of the triangular barrier, and a line of asterisks is constructed as a result of calculations by a numerical method, and a line of squares is used to depict the dependencies for the rectangular barrier, respectively. Therefore, using formula (\ref{as5}), it is possible to quantitatively replace the triangular shape of the barrier with a rectangular one with a width of $h_{1}$.

The field emission current density is determined by the expression:
\begin{equation}
	j=\int_{0}^{\infty}eDNdW,
	\label{as6}
\end{equation}
where $N$ is the number of electrons emit on a metal surface of unit area per unit time with a kinetic energy $W$ normal to the surface. The quantity $N$ was evaluated by Nordheim and under the conditions of the problem is found in the form:
\begin{equation}
	N=\frac{m_{e}}{2\pi^{2}\hbar^{3}}\left( \mu-W\right).
	\label{as7}
\end{equation}
The integrand of formula (\ref{as6}) is the differential current:
\begin{equation}
	dj/dW = eDN.
	\label{as8}
\end{equation}

\begin{figure}
	\includegraphics[width=0.5\textwidth]{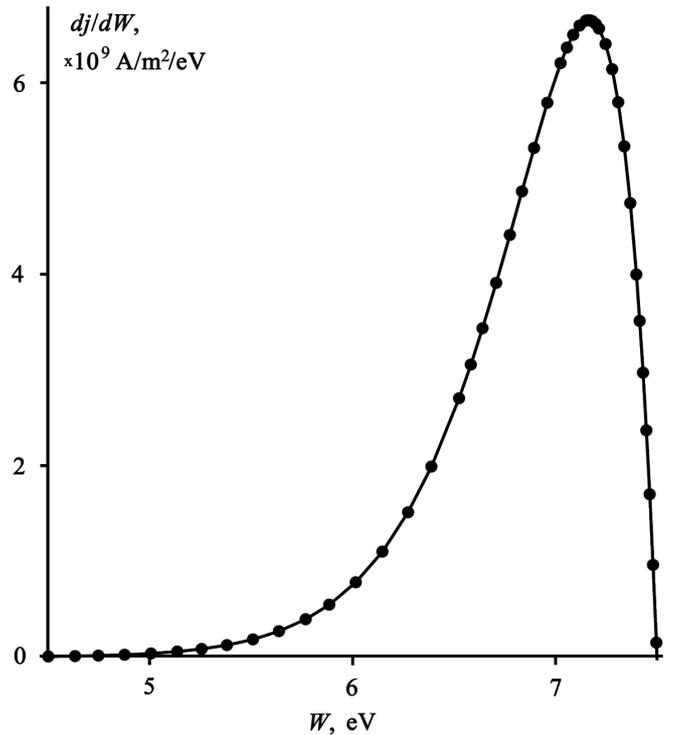}
	\caption{Differential current in the case of single-stage potential barriers: graphs of analytical dependence (solid line) for the triangular shape of the barrier; dependence curve (line of solid circles) in the case of a rectangular barrier}
	\label{penF}       
\end{figure}

The right-hand side of expression (\ref{as8}) determines the electron energy distribution density. Figure~\ref{penF} shows the graphic dependences of the differential current $dj/dW$ on the electron energy at the parameters that were used for Figure~\ref{penE}. In Figure 6 the solid line corresponds to the case of the triangular potential barrier. The line of solid circles is used if the rectangular barrier is considered. As can be seen from Figure~\ref{penF}, the plots of dependences have coincided well for rectangular and triangular shapes of potential barriers.

\section{Analysis of the properties of a double potential barrier of rectangular shape}

The idea of this model is to emit electrons from the metal surface to a vacuum sequentially through two rectangular potential barriers due to the tunneling effect. In the case of the selected parameters, the probability of an electron with energy $W$ passing through a double potential barrier can be higher than for a single-stage barrier, and as a consequence, resonant tunneling of electrons will occur \cite{r25}.

\begin{figure}
	\includegraphics[width=0.5\textwidth]{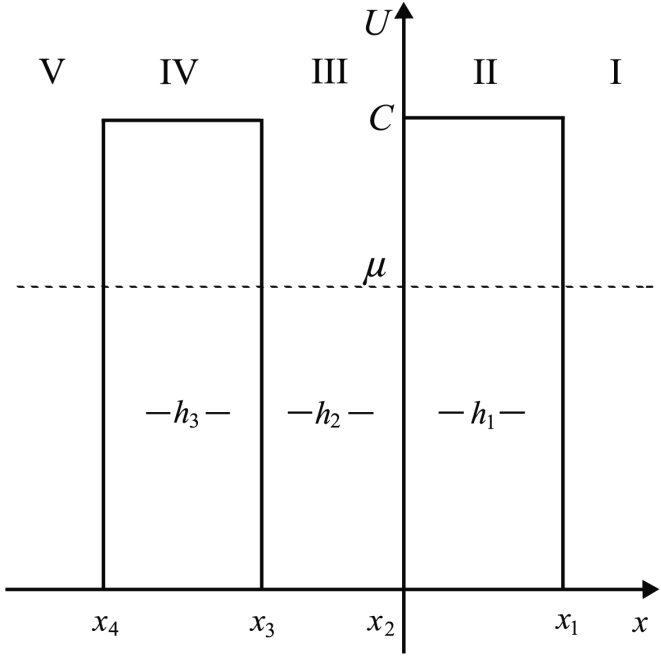}
	\caption{The double potential barrier of rectangular shape}
	\label{penG}       
\end{figure}

Figure~\ref{penG} shows a simplified for computation potential barrier shape for the metal-metal-vacuum medium with a thickness of the nanoscale metal coating, which is used to find the analytical expression of the transmission coefficient and transfer of the revealed mathematical regularities to the real model of the double potential barrier of a system of two layers of metals. The model is similar to the barrier shape to describe a quantum resonant-tunneling diode \cite{r25}. Each of the regions means: V is a metal substrate, IV is a dipole layer of metal-metal contact, III is a coating layer identical to the substrate material, II is a potential barrier at the metal-vacuum interface, I is a vacuum. Region II is a quantitative replacement of triangular barrier by a rectangular one with equal transparency. The $x_{i}$ coordinates are the boundaries of the specified regions. The parameters $h_{1}$, $h_{2}$, $h_{3}$ of the model are the width of regions II, III, IV, respectively. The subscript numbering of the wave vectors $k_{i}$ coincides with the number of each region. If $h_{1} = 0$ or $h_{3} = 0$, the double potential barrier of a rectangular shape becomes single-stage.

The transmission coefficient of the double potential barrier of rectangular shape in Figure~\ref{penG} with the same potential energy $C$ in regions II and IV is found in analytical form:
\begin{equation}
	D_{2}  = \left(1+\frac{\eta^{2}}{4}sh^{2}\left(k_{2}\left(h_{3}+h_{1}\right)\right)+Z\right)^{-1},
	\label{as9}
\end{equation}
where $Z=L\left(\frac{\eta^{4}}{16}L\sin^{2}{\left(k_{1}h_{2}\right)}-P+R\right)$; \\ $L=ch{\left(k_{2}\left(h_{3}+h_{1}\right)\right)}-ch{\left(k_{2}\left(h_{3}-h_{1}\right)\right)}$; \\
$P=\frac{\eta^{2}}{2}\sin^{2}{\left(k_{1}h_{2}\right)}\cdot ch{\left(k_{2}\left(h_{3}+h_{1}\right)\right)}$; \\
$R=\frac{\eta^{3}\left(\varphi-W\right)}{8C}\sin{\left(2k_{1}h_{2}\right)}\cdot sh{\left(k_{2}\left(h_{3}+h_{1}\right)\right)}$; \\
$k_{1}=k\sqrt{W}$; $k_{1}=k_{3}=k_{5}$;
$k_{2}=k_{4}$. \\
In general, the transmission coefficient of a double barrier with different heights of potential energy regions was found in \cite{r24,r28}. If $h_{1} = 0$ or $h_{3} = 0$, then formula (\ref{as9}) is simplified to expression (\ref{as1}). Using the condition of maximum manifestation of resonance:
\begin{equation}
	h_{1}=h_{3}=h,
	\label{as10}
\end{equation}
on condition $W = C/2$, which simplifies the calculation, formula (\ref{as9}) can be written as:
\begin{equation}
	D_{2}=\left(1+\cos^{2}{\left(kh_{2}\sqrt{\frac{C}{2}}\right)}\cdot sh^{2}\left(2kh\sqrt{\frac{C}{2}}\right)\right)^{-1}.
	\label{as11}
\end{equation}

The condition $D_{2} = 1$ for the absolute transparency of a double potential barrier of rectangular shape in the case of the expression (\ref{as10}) and at the value $W = C/2$ is obtained in the form:
\begin{equation}
	h_{2}= \frac{\lambda_{B}}{4}\left(2n+1\right), ~n~=~0,~1,~2,\ldots
	\label{as12}
\end{equation}
where $\lambda_{B}={2\pi\hbar}/{\sqrt{m_{e}C}}$ is the de Broglie wavelength of an electron.

At the electron energy $W=6$~eV then $\lambda_{B} =5\cdot10^{-10}$ ~m, and using (\ref{as12}), $h_{2} = 1.25\cdot10^{-10}$ ~m. It is of importance, that the formula (\ref{as12}), found under condition (\ref{as10}) and the value of $W~=~C~/~2$. The maximum possible value of $D$ is also in the case of violation of the condition (\ref{as10}), but less than one (not absolute transparency). Introduction (\ref{as10}) and $W~=~C~/~2$ is necessary to simplify (\ref{as9}) and find the simple mathematical expression (\ref{as11}) for transparency of the double rectangular barrier.

\begin{figure}
	\includegraphics[width=0.5\textwidth]{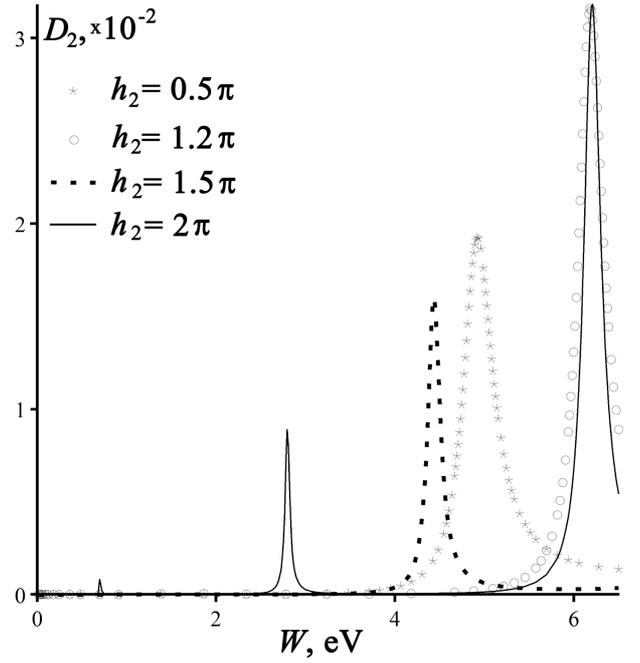}
	\caption{Graphic dependences of the transmission coefficient $D_{2}$ on $W$}
	\label{penH}       
\end{figure}

Applying conditions (\ref{as10}) and $h_{2}=\pi\sqrt{{2}/{C}}\approx$ $1.25\cdot10^{-10}$ ~m, the transmission coefficient (\ref{as9}) can be represented as a second-order Taylor series near the coordinate $W~=~C~/~2$:
\begin{eqnarray}
	D = 1-\frac{4}{C^{2}}\left(ch{\left(h\sqrt{\frac{C}{2}}\right)}+\frac{\pi}{4}sh{\left(h\sqrt{\frac{C}{2}}\right)}-1\right)^{2} \times
	\nonumber\\
	\times {\left(W-\frac{C}{2}\right)^{2}}.
	\label{as13}
\end{eqnarray}
If $h>10^{-10}$ ~m, then formula (\ref{as13}) is simplified to the form:
\begin{equation}
	D=1-\left(\frac{1+\pi/4}{C}\right)^{2}\exp{\left(h\sqrt{2C}\right)}\left(W-\frac{C}{2}\right)^{2}.
	\label{as14}
\end{equation}

Figure~\ref{penH} represents the graphic dependences of the transmission coefficient $D_{2}$ on the electron energy of the double potential barrier from Figure~\ref{penG} at $C = 12$ ~eV, $h_{1} =$  $3\cdot10^{-10}$ ~m, $h_{3} = 10^{-10}$ ~m. As can be seen from Figure~\ref{penH}, the transmission coefficient maxima of the potential barrier are observed for arbitrary $h_{1}$, $h_{3}$ and the height and width of the graphic dependences peaks are increase with increasing energy $W$. If the value of $h_{2}$ increases, the width of the peaks decreases.

\begin{figure}
	\includegraphics[width=0.5\textwidth]{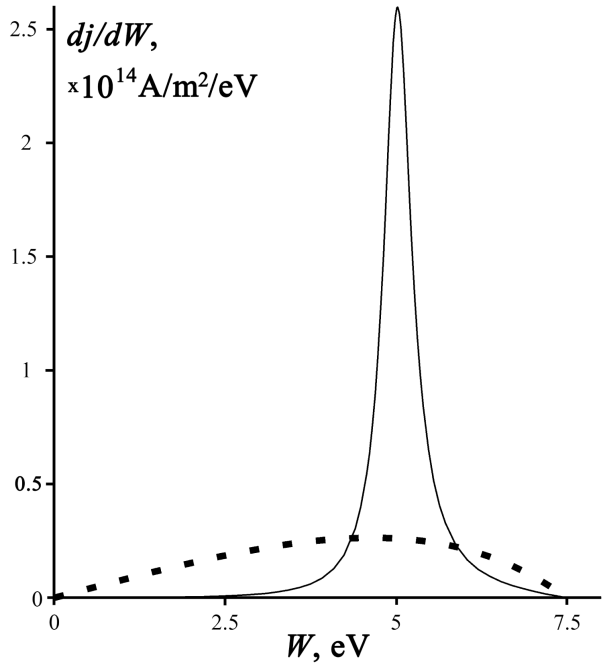}
	\caption{Differential current plots of the single-stage (depicted by dots) and the double (solid line) rectangular potential barriers}
	\label{fig9}       
\end{figure}

Figure~\ref{fig9} illustrates the dependences of the differential current on the energy $W$ under the conditions $C = 12$ ~eV, $\mu = 7.5$ ~eV, $h_{1} = h_{2} = 1.5 \cdot10^{-10}$ ~m, $h_{3} = 10^{-10}$ ~m in the case of a double potential barrier of rectangular shape (the solid line) and rectangular single-stage barrier (the line is depicted by dots). The parameters $h_{1}$, $h_{2}$, $h_{3}$ are selected so that the resonance condition (\ref{as12}) is fulfilled. Thus, the width $h_{2}$ is multiple to a quarter of the de Broglie wave $\lambda_{B}~/~4$, which is similar to a special technique of an anti-reflective coating \cite{r21,r22,r23}. An additional coating applied to the metal substrate not only increases transparency but also reduces the energy scattering of field emission electrons in the region of absolute transparency of the double rectangular potential barrier. This conclusion is confirmed by the graphs in Figure~\ref{fig9}.

\begin{figure}
	\includegraphics[width=0.5\textwidth]{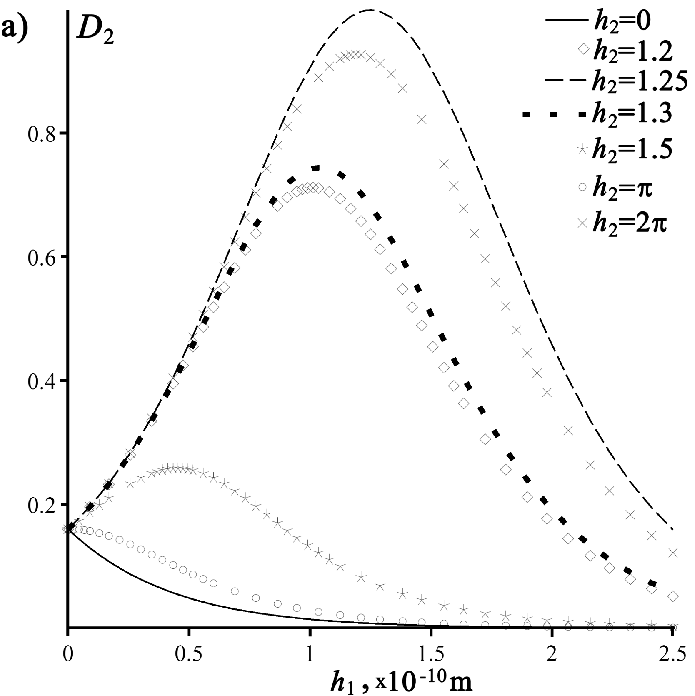}
\end{figure}

\begin{figure}
	\includegraphics[width=0.5\textwidth]{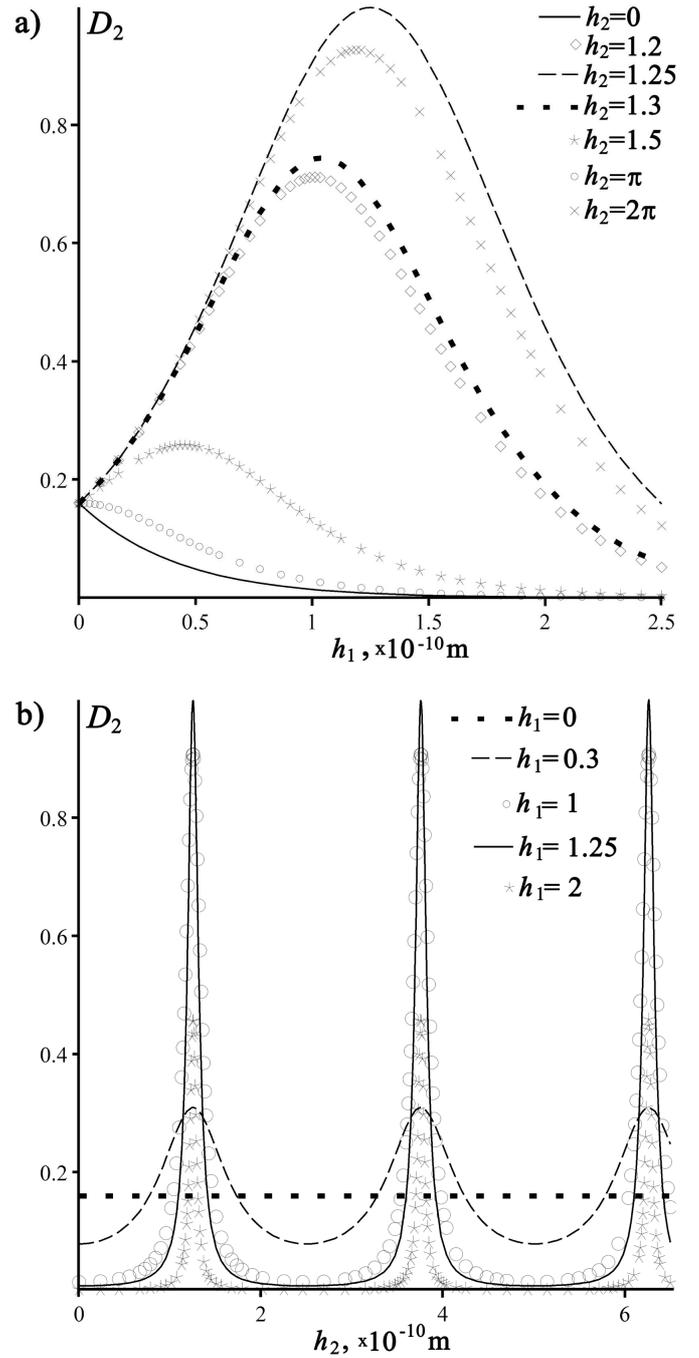}
	\caption{Graphic dependences of the transmission coefficient $D_{2}$ on: a) $h_{1}$; b) $h_{2}$}
	\label{penJ}       
\end{figure}

In Figure~\ref{penJ} a), b) it is seen that if $W = 6$ ~eV, $C = 2W$, $h_{3} = 1.25\cdot10^{-10}$ ~m then the transmission coefficient (\ref{as11}) is the greatest when $h_{2} = \lambda_{B}~/~4$. By increasing the values of $h_{1}$ and $h_{3}$, the transmission coefficient decreases to zero, but the graphic appearance of the dependencies is retained. From Figure~\ref{penJ} b) it can be concluded that there is the repeatability of maxima and minima of $D_{2}$ according to formula (\ref{as12}). These peaks were found numerically because it is difficult to experimentally fulfill the condition for the thickness of the coating on the surface of a metal substrate.

\begin{figure}
	\includegraphics[width=0.5\textwidth]{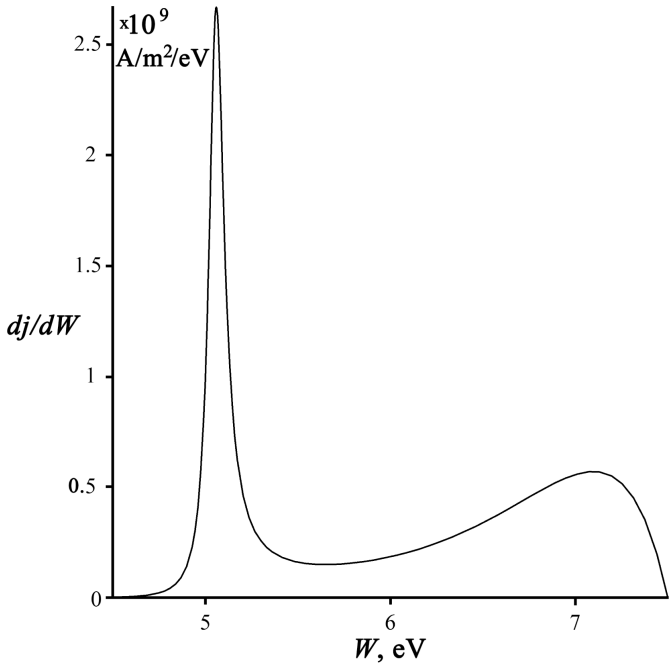}
	\caption{Dependence on $W$ of differential current of the double rectangular potential barrier}
	\label{penK}       
\end{figure}

Figure~\ref{penK} illustrates the graph of the differential current dependence on the electron energy for the barrier model from Figure~\ref{penG}. It is observed the resonance region of the aforementioned quantity near the coordinate of maximum $W_{C} = 5.1$ ~eV. The second peak is the background non- resonance differential current. If there is no potential barrier for the motion of the electron in region IV, which is indicated in Figure~\ref{penG}, the background non-resonance differential current is transformed into a differential field emission current. Figure~\ref{penK} was obtained under conditions: $C = 12$ ~eV, $\mu = 7.5$ ~eV, $h_{2} = h_{3} = 1.5 \cdot10^{-10}$ ~m. The value of $h_{1}$ is determined from expression (\ref{as5}) at $F = 7.5\cdot10^{9}$ ~eV~/~m.

Therefore, for a theoretical study of the field emission phenomenon from a metal in the presence of nanoscale voids in the surface metal layer, the model of rectangular double potential barrier simplified for calculations is proposed.

\section{Simulation of the field emission current through a double barrier in the presence of an external electric field}

For verification with analytical results for the barrier in Figure~\ref{penG}, it is calculated by the numerical method the value of the field emission current density from a two-layer metal system in the case of the potential barrier model from Figure~\ref{penA}. First of all, let us compare the dependences of the transmission coefficient and the differential current on $W$ for the double barrier, which correspond to Figure~\ref{penG}, and the double barrier in Figure~\ref{penA}.

\begin{figure}
	\includegraphics[width=0.5\textwidth]{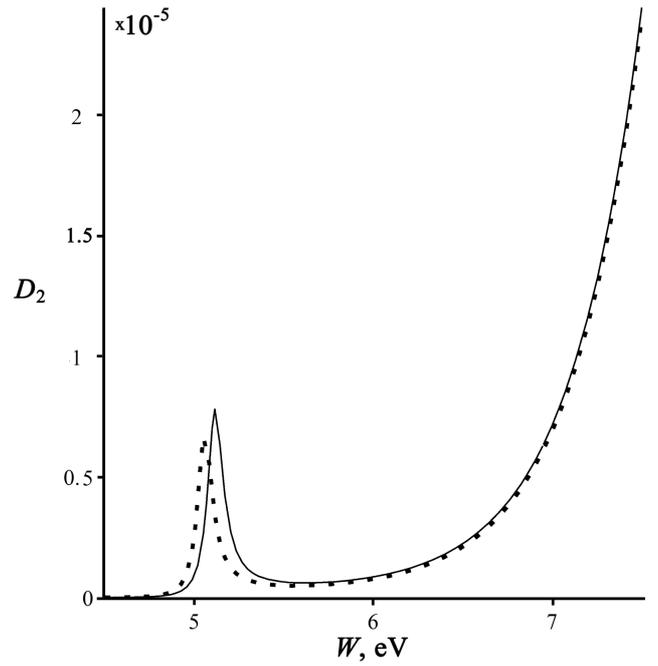}
	\caption{Graphic dependences of $D$ on energy $W$ in cases of double rectangular (the line of dots) and double (the solid line) potential barriers}
	\label{penL}       
\end{figure}

\begin{figure}
	\includegraphics[width=0.5\textwidth]{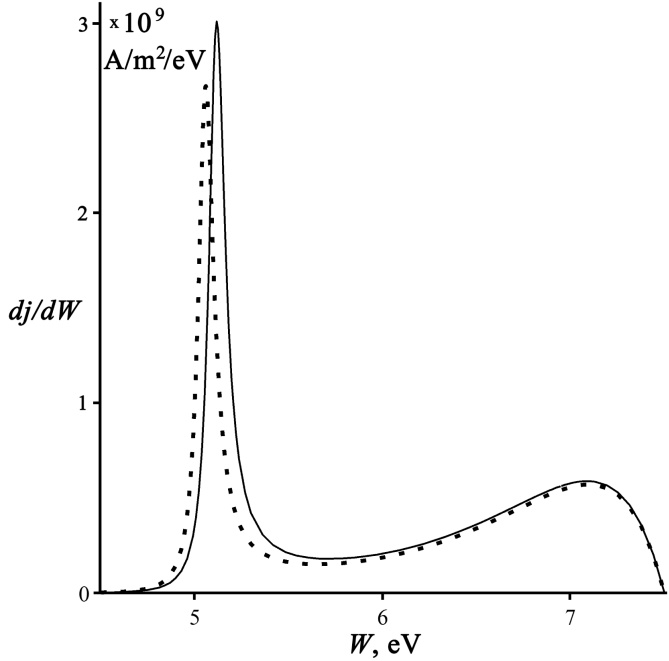}
	\caption{Graphic dependences of differential current on energy $W$ in cases of double rectangular (line of dots) and double (solid line) potential barriers}
	\label{penM}       
\end{figure}

\begin{figure}
	\includegraphics[width=0.5\textwidth]{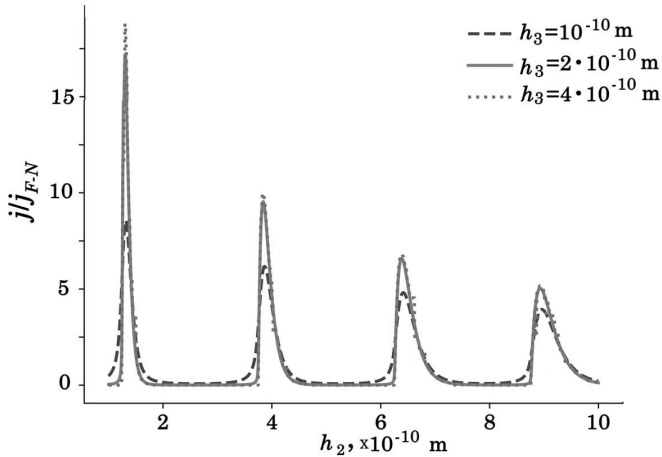}
	\caption{Comparison of the field emission current density $j$ and $j_{F-N}$ in the case of the double potential barrier model}
	\label{penN}       
\end{figure}

Figure~\ref{penL} and Figure~\ref{penM} were obtained under the conditions: $C = 12$ ~eV, $\mu = 7.5$ ~eV, $h_{2} = h_{3} = 1.5 \cdot10^{-10}$ ~m. The value of $h_{1}$ is determined for an arbitrary energy $W$ from expression (\ref{as5}) at $F = 7.5\cdot10^{9}$ ~eV~/~m. The curve depicted by dots corresponds to the dependence for the model in Figure~\ref{penG}. The solid line indicates the dependence graph in the case of the double potential barrier described in Figure~\ref{penA}. Therefore, a quantitative coincidence of dependence graphs is observed at a shift that less than $0.1$ ~eV in the maxima range of physical quantities. Figure~\ref{penN} shows two peaks of the dependence graphs of differential current on the electron energy that will be at other values of $F$.

Figure~\ref{penN} demonstrates the dependence graphs of $j/j_{F-N}$ on $h_{2}$ obtained by the numerical method at different values of $h_{3}$. At the value $E = 5\cdot10^{9}$ ~V~/~m for the copper characteristics: $\mu = 7.5$ ~eV, $\chi = 4.5$ ~eV. Then the current $j_{F-N}$ from the ideal copper surface is $4.2\cdot10^{7}$ ~A~/~m$^{2}$. In Figure~\ref{penN} the maxima of the current density $j$ are more than 5 times greater than $j_{F-N}$ at the above parameters for $h_{2} < 10^{-9}$ ~m.

Properties for the model in Figure~\ref{penG} will be typical for the double potential barrier from Figure~\ref{penA}. Under the conditions of the selected parameters $h_{1}$, $h_{2}$, $h_{3}$ for a system of two metal layers, where the upper layer has a nanoscale thickness, the similar resonant tunneling \cite{r25} of electrons will occur and one can observe a reduction of electron energy scattering in the absolute transparency region. From Figure~\ref{penN} it can be concluded that with increasing $h_{2}$ the current density oscillations are decreasing. But the important contribution to the dark current in accelerating cavities and, hence, to the field emission current density is given by the oscillations of the current even when the value of the parameter $h_{2}$ is more than one nanometer. The greatest value of the current density dependence on $h_{2}$ in Figure~\ref{penN} is the first maximum at the point $h_{2} \approx \lambda_{B}~/~4$, which is similar to a special technique of an anti-reflective coating \cite{r21,r22,r23}.

\section{Width investigation of the resonance oscillatory peak of field emission current}

Let us evaluate the full width at half maximum ($FWHM$) of differential current $dj/dW$ in the resonance region from Figure~\ref{penK}. One can use the expressions (\ref{as13}) and (\ref{as14}) that correspond to the formula (\ref{as9}). One can find the analytical formula for $FWHM$, taking into account the expression (\ref{as14}) and on condition $D(W_{1})~=~0.5$, in the following form:
\begin{equation}
	FWHM = \left( \frac{\sqrt{2}}{1 + \frac{\pi}{4}}\right)C\exp{\left(-h\sqrt{\frac{C}{2}}\right)}.
	\label{as15}
\end{equation}

\begin{figure}
	\includegraphics[width=0.5\textwidth]{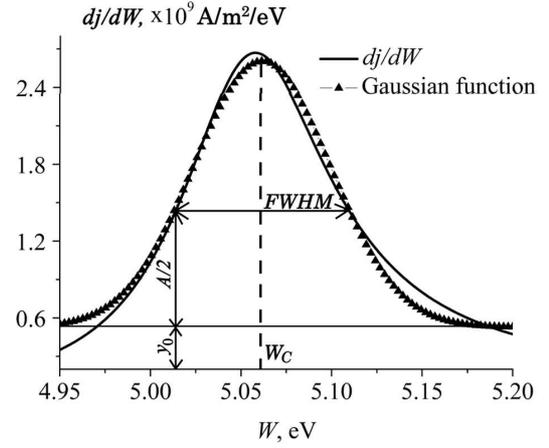}
	\caption{Approximation of the differential current dependence on the electron energy by a Gaussian function in the resonance region}
	\label{penO}       
\end{figure}

The next stage is to approximate the resonant region of the dependence $dj/dW$ from Figure~\ref{penK} by a Gaussian function:
\begin{equation}
	y = y_{0}+A\exp{\left(-\frac{\left(W-W_{C}\right)^2}{2\omega^{2}}\right)}.
	\label{as16}
\end{equation}
Figure~\ref{penO} shows the dependence graph (\ref{as16}), depicted by a line of triangles at numerically found values: $W_{C}~=~5.1$ ~eV, $y_{0}~=~5.3\cdot10^{8}$ ~A~/~m$^2$~/~eV, $\omega~=~0.04$ ~eV, $A~=~2.1\cdot10^{9}$ ~A~/~m$^2$~/~eV. In Figure~\ref{penO} the solid line describes the resonant region of dependence from Figure~\ref{penK}, obtained numerically. The formula for $FWHM$ was found, using (\ref{as16}), and it is written as:
\begin{equation}
	FWHM = 2\omega\sqrt{\ln{4}}.
	\label{as17}
\end{equation}
From formula (\ref{as17}) it is found that $FWHM ~\approx~ 0.1$ ~eV. Thus, the appearance of the resonant maximum region is a characteristic of electrons in a narrow range of energies and selected parameters $h_{1}$, $h_{2}$, $h_{3}$, when the condition (\ref{as12}) is fulfilled.

\begin{figure}
	\includegraphics[width=0.5\textwidth]{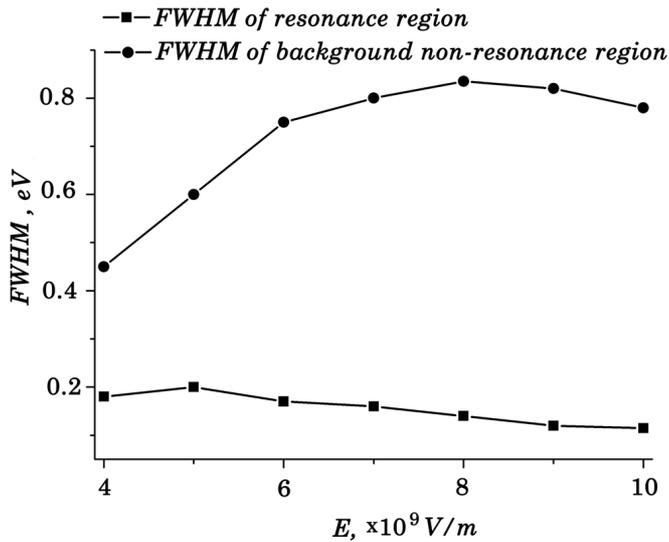}
	\caption{Graphic dependences of $FWHM$ on $E$ in the case of the double potential barrier of rectangular-triangular shape}
	\label{penP}       
\end{figure}

Let us compare the dependence graph of $FWHM$ of the two peaks, constructed by a solid line in Figure~\ref{penM}, at different values of the electric field strength $E$. From Figure~\ref{penP}, constructed under conditions $C = 12$ ~eV, $\mu = 7.5$ ~eV, $h_{2} = h_{3} = 1.5\cdot10^{-10}$ ~m, it is concluded that the $FWHM$ value of the differential current dependence on $W$ from Figure~\ref{penM} in the resonant maximum region at different $E$ is in a narrow range of energies and is equal to $0.1$ ~eV. The value of full width at half maximum in the resonant region of the differential current dependence on $W$ decreases with increasing $h_{3}$. The $FWHM$ values for the background non-resonance differential current peak, shown by the line with circles, were found more than 5 times greater in comparison with the results for the resonance maximum region. The $FWHM$ values, represented by the line with squares in Figure~\ref{penP}, it is found by the numerical method for the barrier model from Figure~\ref{penA}, are consistent with the analytical result (\ref{as17}) in the case of the potential barrier form in Figure~\ref{penG}.

\section{Summary}

A potential barrier model has been proposed that takes into account additional mechanisms of high-gradient high-vacuum breakdowns in accelerating structures. It was experimentally confirmed in reference \cite{r20} that the field emission current from a real metal surface is greater than the value calculated theoretically. It is assumed that the sources of increased dark current in the accelerating cavities and, consequently, the field emission current $j_{F-N}$ from the electrode surface are nanoclusters on the surface and nanoscale voids in the near-surface metal layer.

As a result, it was shown that the field emission current from the metal surface near the nano-objects has an oscillatory resonance feature. The resonance condition (\ref{as12}) determines the size  of the nanocluster on the metal surface or the depth of the nanoscale void from the metal surface whose diameter is a multiple of ${\lambda_{B}}~/~{4}$. The current takes on a maximum value when the transmittance is 1 and provided the size of the nanoscale void $h_{1}
=h_{1eff}~ \approx ~2\varphi~/~(3eE)$.

The full width at half maximum ($FWHM$) of the resonance region does not depend on the electric field strength, but one depends on the height of the potential barrier $C$ and the size $h_3$. At the parameters $C~=~12$ ~eV, $\mu~=~7.5$ ~eV, $h_{2}~=~h_{3}~=~1.5\cdot10^{-10}$~m, $F~=~ 7.5\cdot10^{9}$~eV~/~m, it is found that $FWHM~\approx~0.1$ ~eV.

It is revealed that at $E~=~5$ ~GV~/~m the field emission current density from copper surface increases more than 5 times in the case of double potential barrier model under the resonance condition (\ref{as12}) compared to $j_{F-N}$.

\begin{acknowledgements}
	
	The publication is based on the research provided by the grant support of the National Academy of Sciences of Ukraine (NASU) for research laboratories/groups of young scientists of the National Academy of Sciences of Ukraine to research priority areas of development of science and technology in 2021~-~2022 under contract ~No~ 16~/~01~-~2021~(3).
	
	The authors thank to O.P. Novak for helpful discussions on the paper.
\end{acknowledgements}


\section{Authors contributions}
All the authors were involved in the preparation of the manuscript.
All the authors have read and approved the final manuscript.


\end{document}